\documentclass[aps,prl,twocolumn,superscriptaddress,preprintnumbers,nofootinbib]{revtex4-1}
\usepackage{amsmath,amssymb,graphicx,natbib,bm}
\usepackage{enumerate}
\usepackage{hyperref}
\usepackage{breakurl}
\usepackage{color}

\newcommand{\GeV}{\ensuremath{\mathrm{GeV}}}

\newcommand{\deltaf}{\hat \Delta_F}
\newcommand{\deltas}{\hat \Delta_S}
\newcommand{\nn}{\nonumber}
\newcommand{\invfb}{\text{fb}^{-1}}
\newcommand{\sigv}{\langle \sigma v \rangle}

\newcommand{\kgday}{kg~$\cdot$~d}
\newcommand{\keVr}{\ensuremath{\mathrm{keV}_\mathrm{r}}}

\usepackage[table]{xcolor}
\definecolor{lightred}{rgb}{0.8,0.2,0.2}

\definecolor{lightblue}{rgb}{0.2,0.2,0.8}

\begin{document} 

\preprint{EFI-13-21}

\title{Flavored Dark Matter and R-Parity Violation}

\author{Brian Batell}
\affiliation{Department of Physics, University of Chicago, Chicago, IL,
60637}
\affiliation{Enrico Fermi Institute, University of Chicago, Chicago, IL,
60637}

\author{Tongyan Lin}
\affiliation{Department of Physics, University of Chicago, Chicago, IL,
60637}
\affiliation{Kavli Institute for Cosmological Physics,
University of Chicago, Chicago, IL, 60637}

\author{Lian-Tao Wang}
\affiliation{Department of Physics, University of Chicago, Chicago, IL,
60637}
\affiliation{Enrico Fermi Institute, University of Chicago, Chicago, IL,
60637}
\affiliation{Kavli Institute for Cosmological Physics,
University of Chicago, Chicago, IL, 60637}

\begin{abstract}
Minimal Flavor Violation offers an alternative symmetry rationale to
R-parity conservation for the suppression of proton decay in
supersymmetric extensions of the Standard Model. The naturalness of
such theories is generically under less tension from LHC searches than
R-parity conserving models.  The flavor symmetry can also guarantee
the stability of dark matter if it carries flavor quantum numbers.  We
outline general features of supersymmetric flavored dark matter (SFDM) models within the
framework of MFV SUSY. A simple model of top flavored dark matter is
presented. If the dark matter is a thermal relic, then nearly the
entire parameter space of the model is testable by upcoming direct
detection and LHC searches.
\end{abstract}

\date{\today}

\maketitle

The hierarchy problem and the WIMP miracle independently motivate new
dynamics at the weak scale. It is therefore a compelling possibility
that both the naturalness and dark matter (DM) problems are resolved
by the same new physics. Indeed, the paradigm for physics beyond the
Standard Model (SM) for nearly three decades, weak-scale supersymmetry
(SUSY) with R-parity conservation (RPC), accomplishes precisely this
feat.

However, SUSY with RPC as a natural solution to the hierarchy problem
is facing increasingly stringent constraints from a swath of searches
at the LHC experiments.  For example, searches for multi-jets and
missing momentum have placed limits on gluinos and squarks in the TeV
range~\cite{ATLAS-CONF-2013-047,CMS-PAS-SUS-13-012}.  A characteristic
feature of these searches involves the selection of events with large
missing transverse momentum, as would inevitably occur from the decay
of superpartners to the stable LSP. Thus, a simple way to evade these
constraints is to allow interactions that violate
R-parity~\cite{Hall:1983id,Ross:1984yg,Barger:1989rk,Dreiner:1997uz,Barbier:2004ez}.
With R-parity violation (RPV) the LSP is unstable, resulting in high
multiplicity signatures with leptons and/or jets in the final
states. In particular, final states with jets are generally very
difficult to constrain due to the large QCD background. For recent
studies
see~\cite{Brust:2011tb,Allanach:2012vj,Brust:2012uf,Evans:2012bf,Han:2012cu,Franceschini:2012za,Durieux:2013uqa,Duggan:2013yna}.

While RPV certainly helps to hide SUSY at the LHC, it is a step
backwards both from theoretical and phenomenological perspectives. An
understanding of the stability of the proton in terms of a protective
symmetry is abandoned. Indeed, not all RPV couplings can be allowed
simultaneously as then both baryon- and lepton-number-violating
couplings will mediate rapid proton decay. This has motivated a number
of theoretical proposals to explain the relative size of RPV couplings
from a symmetry
rationale~\cite{Ibanez:1991pr,Dreiner:2005rd,FileviezPerez:2012mj,Graham:2012th,Dreiner:2012ae,Ruderman:2012jd,Frugiuele:2012pe,Krnjaic:2012aj,Franceschini:2013ne,Csaki:2013we,Florez:2013mxa,Krnjaic:2013eta,Monteux:2013mna}.

A particularly interesting proposal to understand the stability of the
proton in SUSY with RPV is to invoke the principle of Minimal Flavor
Violation (MFV)~\cite{Nikolidakis:2007fc,Smith:2008ju,Csaki:2011ge},
in which one assumes that the non-abelian flavor symmetry $G_F =
SU(3)_Q \times SU(3)_{u}\times SU(3)_{d}\times SU(3)_{L}\times
SU(3)_{e}$ is broken only by the Yukawa
interactions~\cite{Chivukula:1987py,Hall:1990ac,Buras:2000dm,D'Ambrosio:2002ex}.
Even in RPC theories, an MFV structure is generally imposed on the
soft SUSY breaking interactions in order to suppress unwanted flavor
changing neutral currents (FCNCs). It has been shown that MFV by itself
suppresses RPV couplings enough to explain the stability of the
proton. Furthermore, the leading RPV superpotential operator relevant
for collider phenomenology is $\bar u \bar d \bar d$, leading to
signatures with multiple jets, bottom quarks and top quarks, which
have much weaker direct LHC constraints compared to RPC SUSY.

While theoretical progress has been made towards understanding the
size of RPV couplings, it would still seem we have forsaken our second
compelling hint for physics at the weak scale -- the WIMP miracle.  As
RPV renders the LSP unstable there is no viable DM candidate amongst
the superpartners of the SM particles.

In this paper we demonstrate that the MFV hypothesis can also provide
a symmetry rationale for WIMP DM.  In Ref.~\cite{Batell:2011tc} it was
shown using an operator analysis that MFV automatically implies exact
stability for a large number of representations of the quark flavor
group $G_q = SU(3)_Q \times SU(3)_{u}\times SU(3)_{d}$, leading to the
scenario of flavored dark matter, where DM is charged under $G_q$. Here we demonstrate that this
stability is the result of an underlying $Z_3$ symmetry, which we term
{\it flavor triality}, that is a subgroup of $SU(3)_c \times SU(3)_Q
\times SU(3)_{u}\times SU(3)_{d}$.  Under this $Z_3$ symmetry the SM
fields and Yukawa spurions transform trivially, while the FDM
candidate is charged.  In Ref.~\cite{Batell:2011tc},
non-supersymmetric theories were investigated.  Here we consider
supersymmetric theories of flavored dark matter (SFDM).  We will
examine the general structure of SFDM models including the effects
of SUSY breaking on the flavor splittings in the mass spectrum and
couplings. 

Finally, we will investigate in detail a model of top flavored dark
matter. The DM candidate is taken to be a vector-like fermion
contained in a gauge singlet, $SU(3)_{u_R}$ flavor triplet.  
A flavor singlet mediator field with SM gauge quantum
numbers of right handed top allows the DM to interact with the SM. The
DM is a thermal relic due to its efficient annihilation to $t \bar t$
pairs in the early universe. By virtue of its coupling to the top, the
DM obtains a sizable one loop coupling to the $Z$ boson, which
mediates spin independent scattering with nuclei at rates that will be
tested by LUX and future ton scale direct detection experiments.
Furthermore, the mediator fields, being colored, can be produced at
the LHC and decay to DM, leaving signatures with missing energy, jets
and tops. For other studies of flavored DM, see
Refs.~\cite{Batell:2011tc,MarchRussell:2009aq,Kile:2011mn,Kamenik:2011nb,Agrawal:2011ze,Kumar:2013hfa,Lopez-Honorez:2013wla}.

\section{R-parity violation and MFV SUSY}
\label{sec:MFVSUSY}

In the minimal supersymmetric standard model (MSSM), the most general
superpotential consistent with gauge invariance and renormalizability
is given by
\begin{eqnarray}
W & = & 
 Y_e L H_d \bar e + Y_u Q H_u \bar u + Y_d Q H_d \bar d + \mu H_u H_d   \nonumber \\  
 & + & \lambda L L \bar e + \lambda'  L Q \bar d  + \lambda''  \bar u \bar d \bar d +
 \mu'  L H_u.
 \label{W}
\end{eqnarray}
It is useful to assign a matter parity to the superfields, $P_M =
(-1)^{3(B-L)}$, with $B$, $L$ the baryon and lepton number,
respectively. Under matter parity the quark and lepton superfields
have charge $-1$ and the Higgs superfields have charge +1. R-parity is
then defined on the spin $s$ component fields as $P_R = (-1)^{2s}
P_M$, under which all of the SM fields are even and the superpartners
are odd.  The terms on the first line of Eq.~(\ref{W}) conserve matter
parity, while those on the second line do not.

The size of the RPV couplings, $\lambda$, $\lambda'$, $\lambda''$,
$\mu'$ is severely constrained by the non-observation of proton
decay. For example, suppose one desired a small RPV $\bar u \bar d
\bar d$ coupling $\lambda''$ in order to facilitate the decay of the
LSP to jets, with the aim of suppressing the amount of missing
energy. Such a coupling must be at least larger than $\sim {\cal
  O}(10^{-8})$ to mediate a decay on detector scales. If there are
also, {\it e.g.,} $LQ \bar d$ couplings $\lambda'$ present at some level, then
squark exchange will mediate the decay of the proton with a lifetime
\begin{equation}
\tau_p \sim 10^{33} \, {\rm yr} \, 
\left( \frac{10^{-19} }{\lambda'} \right)^2 
\left( \frac{10^{-8} }{\lambda''} \right)^2 
\left( \frac{m_{\tilde q} }{ \rm TeV } \right)^4.
\end{equation} 
We see that to get interesting modifications to SUSY collider
signatures from $\bar u \bar d \bar d$, we must require $LQ \bar d$ to
be extremely suppressed to avoid rapid proton decay. This is another
way of saying that one can have $B$ or $L$ violation, but not both.
Besides proton decay, there are a variety of additional strong
constraints on RPV couplings; for a review see
Ref.~\cite{Barbier:2004ez}.

The problem is even worse than this -- one could imagine for example
that sizable RPV couplings exist only amongst the second and third
generation. However, as a result of electroweak symmetry breaking
rotations from the gauge to the mass basis will generally induce
sizable coupling amongst the first two generations. Clearly, it is
therefore desirable to have a symmetry explanation for the suppression
of dangerous RPV couplings.

Minimal flavor violation provides one such symmetry principle to
explain the smallness of unwanted RPV couplings. The MFV hypothesis
promotes the Yukawa couplings to spurion fields transforming under the
nonabelian flavor symmetry $G_F = SU(3)_Q \times SU(3)_{u}\times
SU(3)_{d}\times SU(3)_{L}\times SU(3)_{e}$. It is assumed that these
spurions are the only source of $G_F$ breaking. This assumption
typically is imposed in any case of the soft breaking squark masses and
trilinear scalar couplings to suppress FCNCs. Since the RPV couplings
in the superpotential contain the quark and lepton fields, it is then
natural to ask what constraints MFV places on the size of RPV
couplings. This question has been addressed in
Refs.~\cite{Nikolidakis:2007fc,Smith:2008ju,Csaki:2011ge} which have
shown that MFV suppresses RPV couplings enough to explain the
stability of the proton, while generating neutrino masses and being
consistent with $n-\bar n$ oscillation and dinucleon decay
constraints.

The primary difference in approach
between~\cite{Nikolidakis:2007fc,Smith:2008ju} and \cite{Csaki:2011ge}
is that the latter imposes holomorphicity on the Yukawa spurions,
which is required if one imagines the Yukawas to arise from the vacuum
expectation values of chiral superfields in a UV completion. This
assumption drastically reduces the number of allowed couplings in the
superpotential, thus leading to a more predictive setup. For the
remainder of this paper, we follow~\cite{Csaki:2011ge} and impose
holomorphy on the Yukawa spurions.

In fact, in the limit of massless neutrinos, MFV allows only
one operator in the superpotential~\cite{Csaki:2011ge}:
\begin{eqnarray}
W & = & \frac{1}{2} w'' (Y_u \bar u) (Y_d \bar d) (Y_d \bar d) \equiv  \frac{1}{2} \lambda'' \bar u \bar d \bar d,
\label{Wbaryon}
\end{eqnarray}
 where the effective $\bar u \bar d \bar d$ coupling is given by
 \begin{equation}
 \lambda''_{ijk} = w'' y_i^{(u)}  y_j^{(d)} y_k^{(d)} \epsilon_{jkl} V^*_{il} .
 \end{equation}
One clearly observes the strong Yukawa and CKM suppression in the
effective couplings $\lambda''$ amongst the first and second
generation fields. At the same time, the coupling $\lambda''_{tsb}$,
though also suppressed, is still large enough to cause the LSP to
decay within the LHC detectors. For example, with an ${\cal O}(1)$ MFV
coupling $w''$, one obtains $\lambda''_{tsb} \sim 10^{-4}$.
  
With massless neutrinos, the baryon number violating operator in
Eq.~(\ref{Wbaryon}) is the only superpotential coupling
allowed. Additional lepton number violating couplings could in
principle arise from the K$\ddot {\rm a}$hler potential which has no
constraints from holomorphicity. However, only $\Delta L = 3$
operators are in fact allowed due to an accidental $Z_3^L \subset
SU(3)_L\times SU(3)_e$ symmetry present in the theory. Such operators
can only occur at the nonrenormalizable level, and are
suppressed. Thus the proton is safely stable in the limit of massless
neutrinos. It is straightforward to incorporate neutrino masses into
the theory, which allows for additional sources of lepton number
violation, but MFV still can safely suppress proton decay and be
consistent with various other RPV constraints~\cite{Csaki:2011ge}.

The collider signatures of MFV SUSY depend primarily on the nature of
the LSP. The third generation squarks, due to their large Yukawa
couplings, can easily be split from those of the first and second
generation. It is therefore possible that the stop or the sbottom is
the LSP. A stop LSP will decay due to the interaction in
Eq.~(\ref{Wbaryon}) via $\tilde t  \rightarrow  s  b$, while a sbottom LSP
can decay similarly via $ \tilde b \rightarrow s t$. These decays are
prompt (except perhaps for the sbottom at low $\tan\beta$). 
Other possible LSPs, such as neutralinos, charginos,
or gluinos can easily leave displaced vertices, as they will decay, 
{\it e.g.,} through an offshell stop or sbottom.  Such signatures are very
distinctive, although presently the limits are not very strong. 
For further studies of the collider phenomenology in MFV SUSY, see 
Refs.~\cite{Csaki:2011ge,Berger:2012mm,Berger:2013sir}.

\section{Dark matter stability from MFV}
\label{sec:stability}

We have seen that MFV SUSY provides an attractive framework to explain
the proton stability in supersymmetric theories with RPV. By design,
the LSP will decay with a very short lifetime such that SUSY collider
events have suppressed missing energy. Therefore, there is no WIMP
DM candidate amongst the superpartners of the SM
particles. This is of course true of any scenario with RPV couplings
large enough to hide SUSY at the LHC.  Therefore, one must look
beyond the RPV MSSM if one takes the WIMP miracle
seriously\footnote{Another possible DM candidate outside the WIMP
  paradigm is a light gravitino, which we do not investigate here}.

Despite the fact that MFV SUSY does not contain a WIMP candidate, MFV
itself provides a motivation for DM, as we now discuss. In
Ref.~\cite{Batell:2011tc}, it was shown that the MFV hypothesis implies
absolute stability for certain representations that transform
nontrivially under the quark flavor group $G_q = SU(3)_Q \times
SU(3)_{u}\times SU(3)_{d}$ and are singlets under $SU(3)_c$. If these
states are also electrically neutral, they make excellent dark
matter candidates.

In Ref.~\cite{Batell:2011tc} DM stability was demonstrated through
an operator analysis. Conditions were derived for the existence of the
most general operator composed of a single DM multiplet, SM
fields and Yukawa spurions that would mediate the decay of the
would-be DM particle if present.  Here we wish to revisit
this question from a symmetry perspective. We will show that DM
stability can be traced to the presence of an accidental $Z_3$
symmetry, which we call {\it flavor triality}, that is present under
the assumption of MFV.

It is in fact very simple to see that an accidental symmetry is
present that can stabilize DM.  Consider the following discrete $Z_3$
transformation which is an element of $SU(3)_c \times SU(3)_Q \times
SU(3)_{u} \times SU(3)_{d}$ :
\begin{equation}
U = \left(\omega^2 \right)_{c}  \times  \left(\omega \right)_{Q} \times \left(\omega \right)_{u}  \times \left(\omega \right)_{d},
  \label{transformationA}
\end{equation}
where $\omega \equiv e^{2\pi i/3}$ and the subscript indicates the
group which contains the corresponding $Z_3$ element.  Using the
representations listed in Table~\ref{tab:reps}, one can easily check
that {\it all} of the SM fields and Yukawa spurions transform
trivially under (\ref{transformationA}). For example, under
(\ref{transformationA}) $Q \rightarrow \omega^3 Q = Q$, $\bar u
\rightarrow \omega^{-3} \bar u = \bar u$, etc.

%
\begin{table}
\begin{center}
\renewcommand{\arraystretch}{1.3}
 \begin{tabular}{| c || c | c | c | c|}
\hline
   & ~~$SU(3)_c$~
    &  $SU(3)_Q$   &$SU(3)_{u}$   & $SU(3)_{d}$  \\
 \hline
 $Q$ & $\mathbf{3}$   &  $\mathbf{3}$ & $\mathbf{1}$ & $\mathbf{1}$ \\ \hline
$\bar u$ & $\mathbf{\bar 3}$   &  $\mathbf{1}$ & $\mathbf{\bar 3}$ & $\mathbf{1}$ \\ \hline
$\bar d$ & $\mathbf{\bar 3}$   &  $\mathbf{1}$ & $\mathbf{\bar 3}$ & $\mathbf{1}$ \\ \hline
$Y_u$ & $\mathbf{1}$   &  $\mathbf{\bar 3}$ & $\mathbf{3}$ & $\mathbf{1}$ \\ \hline
$Y_d$ & $\mathbf{1}$   &  $\mathbf{\bar 3}$ & $\mathbf{1}$ & $\mathbf{3}$ \\ \hline
$G$ & $\mathbf{8}$   &  $\mathbf{1}$ &   $\mathbf{1}$ &   $\mathbf{1}$ \\ \hline
\hline
  \end{tabular}
\end{center}
\caption{
Representations of fields charged under 
$SU(3)_c \times SU(3)_Q \times SU(3)_{u} \times SU(3)_{d}$  
}
\label{tab:reps}
\end{table}

Now consider a new matter multiplet $\chi$ which is to be our DM
candidate.  This field is a color singlet (as DM is both color and
electrically neutral) and transforms under $G_q$ with irreducible
representation
\begin{equation}
\chi  \sim (n_Q, m_Q)_Q \times  (n_u, m_u)_u \times  (n_d, m_d)_d,
\label{chirep}
\end{equation}
where we have used tensorial notation with $n_Q, m_Q$ etc. taking
possible values $0,1,2,\dots$. Under the $Z_3$ transformation
(\ref{transformationA}), $\chi$ transforms as
\begin{equation}
\chi \rightarrow \omega^{n-m} \chi, 
\label{chitransform}
\end{equation}
where we have defined $n \equiv n_Q + n_u + n_d$, $m \equiv m_Q + m_u
+ m_d$. Thus, $\chi$ will transform nontrivially under
(\ref{transformationA}) provided the following condition is met:
\begin{equation}
(n - m) {\rm mod} \,3 \neq 0.
\label{eq:cond}
\end{equation}
Since the SM fields and Yukawa spurions transform trivially under the
$Z_3$ transformation, provided the condition
(\ref{eq:cond}) is met, $\chi$ transforms nontrivially and will
therefore be absolutely stable. We will use the term {\it flavor
  triality} to refer to this $Z_3$ symmetry under which DM is charged
and the SM is neutral.  Examples of $G_q$ representations for which
(\ref{eq:cond}) holds are flavor triplets, ({\it e.g.,}
$(\mathbf{3},\mathbf{1},\mathbf{1})$ etc.), sextets, ({\it e.g.,}
$(\mathbf{6},\mathbf{1},\mathbf{1})$ etc.) and certain mixed
representations, ({\it e.g.,} $(\mathbf{3},\mathbf{3},\mathbf{1})$
etc.). On the other hand flavor singlets, octets and certain mixed
representations, ({\it e.g.,} $(\mathbf{3},\mathbf{\bar 3},\mathbf{1})$
etc.)  do not meet \ref{eq:cond}. We wish to emphasize that flavor triality 
does not require SUSY; it is simply a consequence of MFV.

Provided we strictly enforce MFV, the conclusion is that the dark
matter candidate $\chi$ is exactly stable if (\ref{eq:cond}) holds,
even in the presence of arbitrary higher dimension operators. A
natural question is whether stability holds in the presence of
deviations from MFV. Without reference to a concrete UV completion,
perhaps the most sensible way to parametrize deviations from the MFV
hypothesis is to consider additional spurions that break
flavor. Provided that these new spurions transform trivially under
flavor triality, then DM stability will not be spoiled.  For example,
additional spurions transforming like the SM Yukawas or, {\it e.g.,} as
flavor octets will not cause the decay of the DM.

The SM flavor symmetries are anomalous, so one may worry that flavor
triality is broken at the quantum level. An attractive possibility is
that the SM flavor symmetries are gauged in the UV, which ultimately
implies the presence of additional matter that would render the theory
anomaly free. Flavor triality would then fundamentally be a discrete
gauge symmetry and could not be spoiled by quantum gravitational
effects~\cite{Krauss:1988zc}.  This possibility is similar in spirit
to obtaining R-parity from broken gauged
$U(1)_{B-L}$~\cite{Mohapatra:1986su,Martin:1992mq}.

Finally, we wish to comment on the possibility of lepton-flavored dark matter. 
Recall that the stability of quark-flavored dark matter considered above rests on the fact that 
the quark fields transform trivially under the element (\ref{transformationA})
of $SU(3)_c \times SU(3)_Q \times SU(3)_{u_R} \times SU(3)_{d_R}$. 
This is a consequence of the $SU(3)_c$ charge of the quark fields. 
However, since the leptons are not colored, there is no transformation analogous to (\ref{transformationA})
under which the lepton fields transform trivially. Therefore, MFV does not in general imply stability for 
lepton-flavored dark matter.

\section{Flavored Dark Matter in MFV SUSY}
\label{sec: FDM}

With the stability condition (\ref{eq:cond}) we are now in a position
to consider explicit models of flavored DM within the framework of MFV
SUSY. With only the addition of a DM multiplet, interactions between
the DM and MSSM fields are not possible at the renormalizable level. One
could consider effective theories of flavored DM, as was done in
Ref.~\cite{Batell:2011tc}, but in such theories one cannot address
important questions relevant for supersymmetric theories such as gauge
coupling unification.  Renormalizable theories require an additional
mediator field that couples DM to the MSSM fields. See
Refs.~\cite{Agrawal:2011ze},\cite{Kumar:2013hfa} for renormalizable FDM models in
non-supersymmetric setups.

The basic models of SFDM contain a vector-like DM multiplet
$X$, ${\overline X}$ and a vector-like mediator field $Y$, ${\overline Y}$. 
The superpotential is given by
\begin{equation}
W = \hat M_X  \, X \,  {\overline X}  + \hat  M_Y\,  Y \, {\overline  Y} + \hat  \lambda \, X \, Y \, \Phi_{\rm SM},
\label{Wsfdm}
\end{equation}
where $\Phi_{\rm SM}$ is one of the quark fields $Q$, $\bar u$, or
$\bar d$. For simplicity we will restrict our discussion to models in
which $X$ is a flavor triplet and SM gauge singlet, while
$Y$ is a flavor singlet and carries SM gauge charges, but of
course other possibilities exist.

One of the main differences between non-supersymmetric theories of FDM
and SFDM is the constraint that holomorphy places on the
superpotential mass and coupling parameters. No further Yukawa
insertions are allowed in the superpotential, which naively leads one
to expect that the masses and couplings of the individual flavors of
$X$ are not split. Obviously this constraint is not present in
non-supersymmetric theories. However, as we will see below,
non-holomorphic terms in the K$\ddot{\rm a}$hler potential can lead to
large flavor splittings, which will be important for phenomenology.

\paragraph{\bf The mediator $Y$}

The mediator $Y$ will generally be charged under the gauge group of
the SM, and will therefore typically ruin gauge coupling
unification. To avoid this, we can embed $Y$, $\overline Y$ in a
complete SU(5) multiplet.  For example, if
$X  \sim (\mathbf{1},\mathbf{1},\mathbf{3})_{G_q}$, 
then one can embed $Y$ into a 
$\mathbf{5}  \supset  (\mathbf{3}, \mathbf{1},-\tfrac{1}{3})_{\rm SM}$
which allows the coupling 
$X\, Y \, \bar d$. Alternatively, if 
$X \sim (\mathbf{\bar 3},\mathbf{1},\mathbf{1})_{G_q}$ or 
$X \sim (\mathbf{1},\mathbf{3},\mathbf{1})_{G_q}$, we can embed $Y$ in a 
$ \mathbf{\overline {10} }  \supset 
(\mathbf{\bar  3}, \mathbf{\bar 2},- \tfrac{1}{6})_{\rm SM} + ( \mathbf{ 3}, \mathbf{1},\tfrac{2}{3})_{\rm SM}$, which allows couplings to $Q$ or $\bar u$, respectively. 
Thus, our framework is compatible with gauge coupling unification.

\paragraph{\bf  Flavor splittings}
\label{sec:soft}

As mentioned above, the masses and couplings of the individual flavors
are degenerate at the level of the renormalizable superpotential due
to holomorphy of the Yukawa spurions. However, there are several ways
by which flavor splittings can be induced as we now discuss.

There is no holomorphy constraint on the K$\ddot{\rm a}$hler
potential, which may contain additional Yukawa insertions that are
consistent with the flavor symmetry. The kinetic term need not be
canonical, {\it e.g.},
\begin{equation}
\label{kinetic}
\int d^4 \theta  \left( 
X^\dag \, {\hat k}_X \, X  +  
 \overline X \,  \hat { \overline k}_X  { \overline X}^\dag  \right),
\end{equation}
where $\hat k_X$, $\hat{ \overline k}_X$ are matrices in flavor space
which contain Yukawa insertions.  For example, if 
$X \sim (\mathbf{1},\mathbf{3},\mathbf{1})_{G_q}$, we have
\begin{eqnarray}
\hat k_X & = &  1 + k  \, Y_u Y_u^\dag + \dots , \nonumber \\
\hat {\overline k}_X & = &   1 + \overline k  \, Y_u Y_u^\dag + \dots, 
\label{kX}
\end{eqnarray}
where we have written explicitly the leading Yukawa insertion. If the
MFV couplings $k$, $\bar k$ are $O(1)$, this can lead to a sizable
splitting between the masses and physical couplings of the third and
first two generations of flavors after canonical normalization of the
kinetic term is carried out.

Another important effect comes from SUSY breaking terms in the
K$\ddot{\rm a}$hler potential, which can generate {\it effective}
non-holomorphic mass terms in the superpotential, {\it e.g.},
\begin{eqnarray}
&&  ~~ \int d^4 \theta \left(
 \frac{S^\dag}{M} { \overline X} \, \hat \mu_X \, X  + {\rm h.c.} \right) 
\nonumber \\  &  =  & ~~ 
 \int d^2 \theta \left(
 \frac{F}{M} { \overline X} \, \hat \mu_X \,  X  + {\rm h.c.} \right), 
 \label{effectivesuper}
\end{eqnarray}
where $S= \theta^2 F $ is a spurion parameterizing SUSY breaking, with
$\sqrt{F}$ the SUSY breaking scale and $M$ the messenger scale. Since
it originates in the K$\ddot{\rm a}$hler potential, the coupling
$\hat \mu_X$ need not be holomorphic, {\it e.g.}, for 
$X \sim (\mathbf{1},\mathbf{3},\mathbf{1})_{G_q}$, we have
\begin{eqnarray}
\hat \mu_X & = &  \mu_0 + \mu_1\, Y_u Y_u^\dag + \dots,
\label{mueff}
\end{eqnarray}
where again we have written explicitly the leading Yukawa insertion
and $\mu_0$, $\mu_1$ are $O(1)$ MFV couplings.

These two effects, Eqs.~(\ref{kinetic}) and (\ref{effectivesuper}),
are enough to generate large splittings in the masses and couplings
between the top flavor and up and charm flavors, in the example
$X \sim (\mathbf{1},\mathbf{3},\mathbf{1})_{G_q}$. 
Typically, up-type Yukawa insertions will lead to a bigger splittings,
although this is not necessarily true at large $\tan\beta$ since the
bottom Yukawa coupling can be $O(1)$. For example, in the case of
$X \sim (\mathbf{1},\mathbf{1},\mathbf{3})_{G_q}$, one can
expect at large $\tan \beta$ sizable splittings between the bottom
flavor and down and strange flavors of $X$.
 
\paragraph{\bf Majorana mass term?}

It may be of phenomenological interest to consider 
the possible existence of a Majorana mass term for the DM, 
$W \supset \delta M_X X X + \overline{\delta M}_X \bar X \bar X$. 
For instance, such a term will split  Dirac fermion DM into two Majorana states, which can dramatically affect the predictions for direct detection rates. 
In the context of flavored dark matter, this possibility was explored in Ref.~\cite{Kumar:2013hfa}, where it was noted that such a Majorana mass violates MFV and can be regarded as an additional flavor-breaking spurion ({\it e.g.}, if $X$ is a flavor triplet then $\delta M_X$ is an flavor anti-sextet). We would like to emphasize that such a spurion is charged under flavor triality (\ref{transformationA}) and will generically spoil the DM stability unless further symmetries are invoked.

\paragraph{\bf  The scale of SFDM}
\label{sec:soft}

The superpotential masses $\hat M_X$, $\hat M_Y$ in Eq.~(\ref{Wsfdm})
can naturally be tied to the weak scale through SUSY breaking, as can
be seen from Eq.~(\ref{effectivesuper}) in which the effective mass
term generated is of order $m_{\rm soft} = F/M$. This is analogous to
the Giudice-Masiero solution to the $\mu$
problem~\cite{Giudice:1988yz}.

\section{Example: Top Flavored Dark Matter}

We now describe the phenomenology of a concrete model of SFDM in which
the DM carries ``top'' flavor The model contains the following
additional chiral multiplets with quantum numbers,
\begin{eqnarray}
  X_i & \supset & (\eta_i, \chi_i) \sim  (\mathbf{1},\mathbf{1},0)_{\rm SM} \times (\mathbf{1},\mathbf{3},\mathbf{1})_{G_q}, \nonumber \\
  Y & \supset &\, (\phi, \,\psi)  \,\sim  (\mathbf{3},\mathbf{1},\tfrac{2}{3})_{\rm SM} \times (\mathbf{1},\mathbf{1},\mathbf{1})_{G_q}, 
\label{TFDM}
\end{eqnarray}
along with conjugate fields $\overline X^i$, $\overline Y$.  The index
$i = u,c,t$ denotes the flavor of $X$, and $\eta_i$ $(\phi)$ and
$\chi_i$ $(\psi)$ are the scalar and fermionic components of $X_i$
$(Y)$, respectively.  The superpotential is given by
\begin{equation}
W = \hat M_X \, X_i \, \overline X^i  +  \hat M_Y \, Y \, \overline Y+ \hat \lambda \,  X_i \, Y \, \bar u^i.
\label{TFDM-superpotential}
\end{equation}
The DM candidate can in principle be either a scalar or
fermion component of $X_i$; for concreteness we will focus on
fermionic DM $\chi_i$. As discussed in the previous section, the
masses and couplings of the individual flavors in $X_i$ are
degenerate at the level of the renormalizable superpotential, which is
a consequence of holomorphy of the Yukawa spurions. However,
non-holomorphic and SUSY breaking terms in the K$\ddot{\rm a}$hler
potential will generically induce flavor splittings which in this
model can easily be $O(1)$ between the first two and third generations
due to the large top Yukawa coupling, as we now discuss.

First, the non-canonical kinetic terms as in
Eqs.~(\ref{kinetic},\ref{kX}) can be brought to canonical form through
field redefinitions, $X \rightarrow Z_X X$. $\overline{X} \rightarrow
\overline Z_X \overline X$, where
\begin{eqnarray}
Z_X \approx  {\rm diag}(1,1,(1+ k  y_t^2)^{-1/2}), \nonumber \\
\overline Z_X  \approx  {\rm diag}(1,1,(1+ \bar k  y_t^2)^{-1/2}),
\label{canonicalN}
\end{eqnarray}
with $k, \bar k$ $O(1)$ MFV couplings defined in Eq.~(\ref{kX}).
Furthermore, accounting for SUSY breaking terms in the K$\ddot{\rm
  a}$hler potential which generate effective non-holomorphic
superpotential mass terms as in
Eqs.~(\ref{effectivesuper},\ref{mueff}), the masses for the fermions
$\chi^T_i = (\chi_u, \chi_c, \chi_t)$ can be written as as
\begin{eqnarray}
{\cal M}_{\chi} &  =  &  \bar Z_X \left(\hat M_X + \frac{F}{M}  \hat \mu_X \right) Z_X  \nonumber  \\
& \approx & {\rm diag} \left(m,m, \frac{m +(F/M) \mu_1 y_t^2 }{\sqrt{(1+ k y_t^2 )   ( 1+ \bar k y_t^2)}}\right), 
\end{eqnarray}
where we have defined $m =\hat M_X + (F/M)\mu_0$, with $\mu_0, \mu_1$
$O(1)$ MFV couplings defined in Eq.~(\ref{mueff}). We observe that the
first two generation dark fermions $\chi_u, \chi_c$ are degenerate
up to fine splittings induced by the up and charm Yukawas, while the
third generation top-flavored fermion $\chi_t$ can obtain a large
mass splitting due to the $O(1)$ top Yukawa. In particular, $\chi_t$
can be the lightest state in the new sector in which case it will be
stable and the DM candidate, and we specialize to this case
for the remainder of the paper.

The main interactions governing the cosmology and phenomenology of the
model come from the superpotential interaction with $\bar u$ in
Eq.~(\ref{TFDM-superpotential}). In component form, the important
terms are
\begin{equation}
-{\cal L} \supset \bar u_R^i\, \lambda_i^j \, \chi_j \, \phi + \tilde  u_R^{\dag i} \, \lambda_i^j \, \chi_j \, \psi + \overline u_R^i \, \lambda_i^j \, \eta_j  \, \psi +{\rm h.c.}, ~~~
\end{equation}
where again $\eta^T_i = (\eta_u,\eta_c, \eta_t)$ are the scalar
components of $X_i$, and $\phi$ ($\psi$) is the scalar
(fermion) component of the mediator field $Y$.  The couplings
$\lambda$ are split due to the canonical normalization of $X$
(\ref{canonicalN}):
\begin{eqnarray}
\lambda & = & \hat \lambda \,Z_X \nonumber \\
 & \approx & {\rm diag} \left(\hat \lambda, \hat \lambda , \frac{\hat \lambda }{\sqrt{(1+ k y_t^2 ) }}\right).
\end{eqnarray}
This demonstrates that $\chi_u$, $\chi_c$ have identical couplings
up to the small $y_u, y_c$ induced splittings, while the DM
particle $\chi_t$ can have a larger or smaller coupling depending on the value of $k$. 
For later reference we write explicitly the terms
involving the DM candidate $\chi_t$,
\begin{equation}
-{\cal L} \supset \lambda_t \, \bar t_R\,  \chi_t \, \phi + \lambda_t \, \tilde  t_R^* \, \chi_t \, \psi +{\rm h.c.}, ~~~
\label{eq:interactions}
\end{equation}
where $\lambda_t \equiv \hat \lambda/\sqrt{1+k y_t^2} $.

\begin{figure}
\begin{center}
\includegraphics[width=0.46\textwidth]{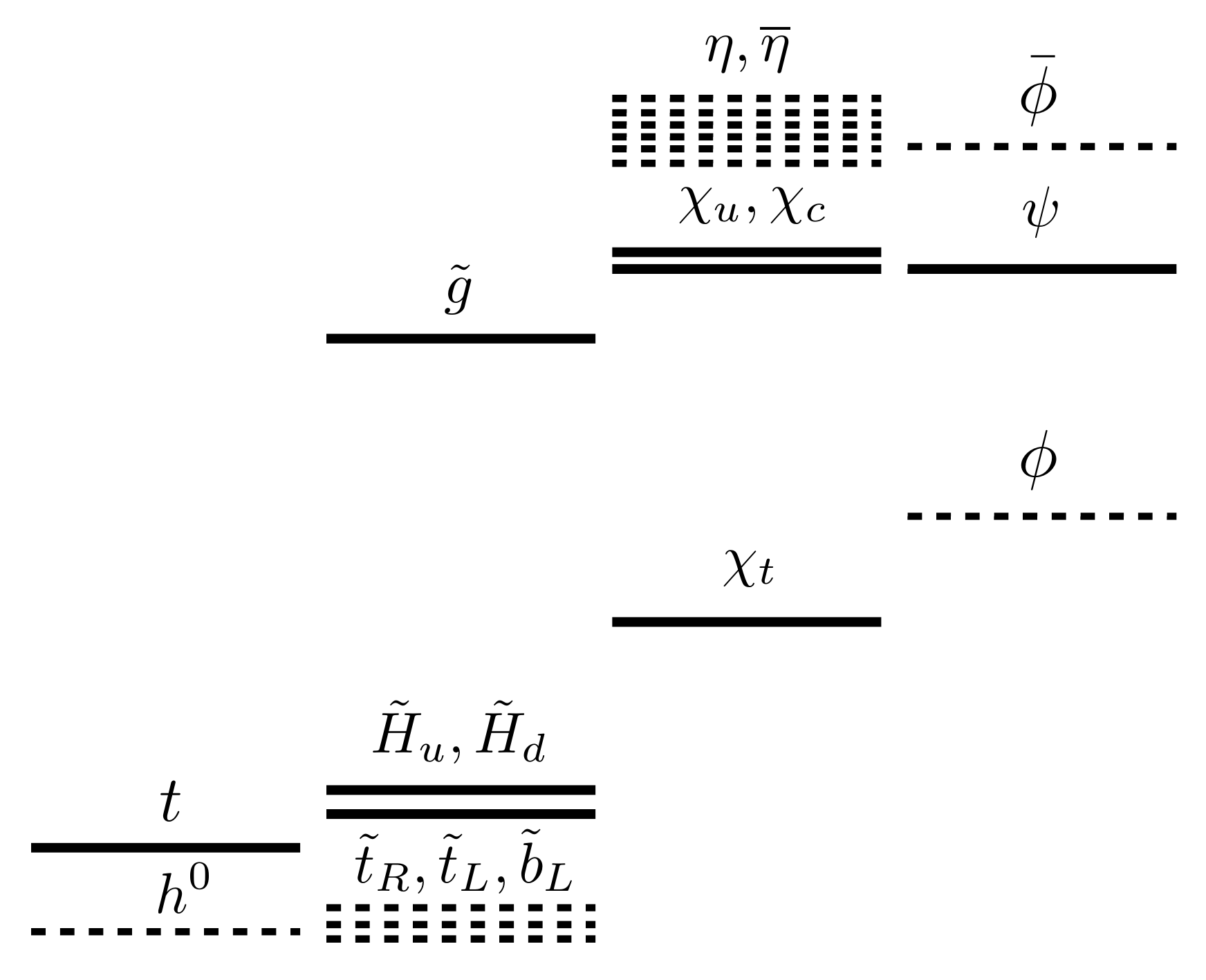}
\caption{Example spectrum of the top-flavored DM model.  The DM is
  $\chi_t$.
    \label{fig:spectrum}}
\end{center}
\end{figure}

Additional soft scalar squared mass terms will split the scalar
components from their fermionic partners. In particular, the
additional scalars $\eta_i$, $\overline \eta_i$ can be raised and will
not be critical to our phenomenological discussion below, although
they can lead to interesting flavor specific signatures if produced at
the LHC. Of more direct importance is the mass of the mediators,
$\phi$, $\bar \phi$ and $\psi$, which play a crucial role in $\chi_t$
annihilation processes in the early universe, mediate the scattering
of $\chi_t$ with nuclei, and by virtue of their $SU(3)_c$ charge can
be directly produced at the LHC. At the supersymmetric level, the
scalar mediators $\phi$ and $\bar \phi$ are degenerate, but SUSY
breaking can split and mass mix these states through soft scalar
squared masses and $b$ terms.  We will make the simplifying
assumptions below that $\phi$-$\bar \phi$ mixing is negligible and
that $m_\phi \ll m_{\bar \phi}, m_\psi$. In this regime $\phi$ will
mediate the dominant interaction between the DM and the SM.  We
display in Fig.~\ref{fig:spectrum} one possible spectrum of the
top-flavored scenario.

We note Ref.~\cite{Kumar:2013hfa} considered a nonsupersymmetric
scenario of top flavored dark matter as an explanation to the Tevatron
top quark forward backward asymmetry. They focused on light DM,
$m_{\chi_t} \lesssim 100$ GeV, whereas we will be concerned with DM
masses $m_{\chi_t} > m_t$.

\paragraph{\bf Thermal relic abundance}

In the regime $m_{\chi_t} > m_t$ and $m_\phi \ll m_\psi$, the dominant
process governing the relic abundance of $\chi_t$ is
\begin{eqnarray}
\chi_t  \bar \chi_t  & \rightarrow  & t  \, \bar t,  \label{anntop} 
\end{eqnarray}
which is mediated by $t$-channel exchange of the scalar mediator $\phi$.
The thermally averaged annihilation cross section is
\begin{eqnarray}
\langle \sigma v \rangle_{t \bar t} & = & \frac{N_c \lambda_t^4 m_{\chi_t}^2 }{32 \pi (m_{\chi_t}^2 + m_{\phi}^2 - m_t^2)^2} 
\left(1 -\frac{m_t^2}{m_{\chi_t}^2} \right)^{1/2}, ~~~~~   \label{anntop} 
\end{eqnarray}
where $N_c = 3$ is the number of colors.  The process $\chi_t \bar
\chi_t \rightarrow \tilde t_i \, \tilde t_j^*$ can also be important
if $\psi$ is similar in mass to $\phi$. While this will result in
small numerical changes to the parameters for which the correct relic
abundance is obtained, it will not qualitatively change the
conclusions presented below.

The observed cold DM relic abundance of $\Omega h^2 \approx 0.12$
\cite{Ade:2013zuv} is obtained for an annihilation cross section of
$\langle \sigma v \rangle~\approx~1.5$~pb (for a Dirac fermion). A
minimum coupling of $\lambda_t = 0.35$ is needed to achieve this cross
section if the $t\bar t$ final state is considered. Exceptions to this
statement are if annihilation to the other final states are present
and substantial, or when $m_\phi - m_{\chi_t} < m_{\chi_t}/20$ and the
effects of coannihilation can assist in achieving the correct relic
density. We do not include the effects of coannihilation in our
numerical results below, but see Ref.~\cite{Bai:2013iqa} for a recent
study.

\begin{figure*}
\begin{center}
\includegraphics[width=0.47\textwidth]{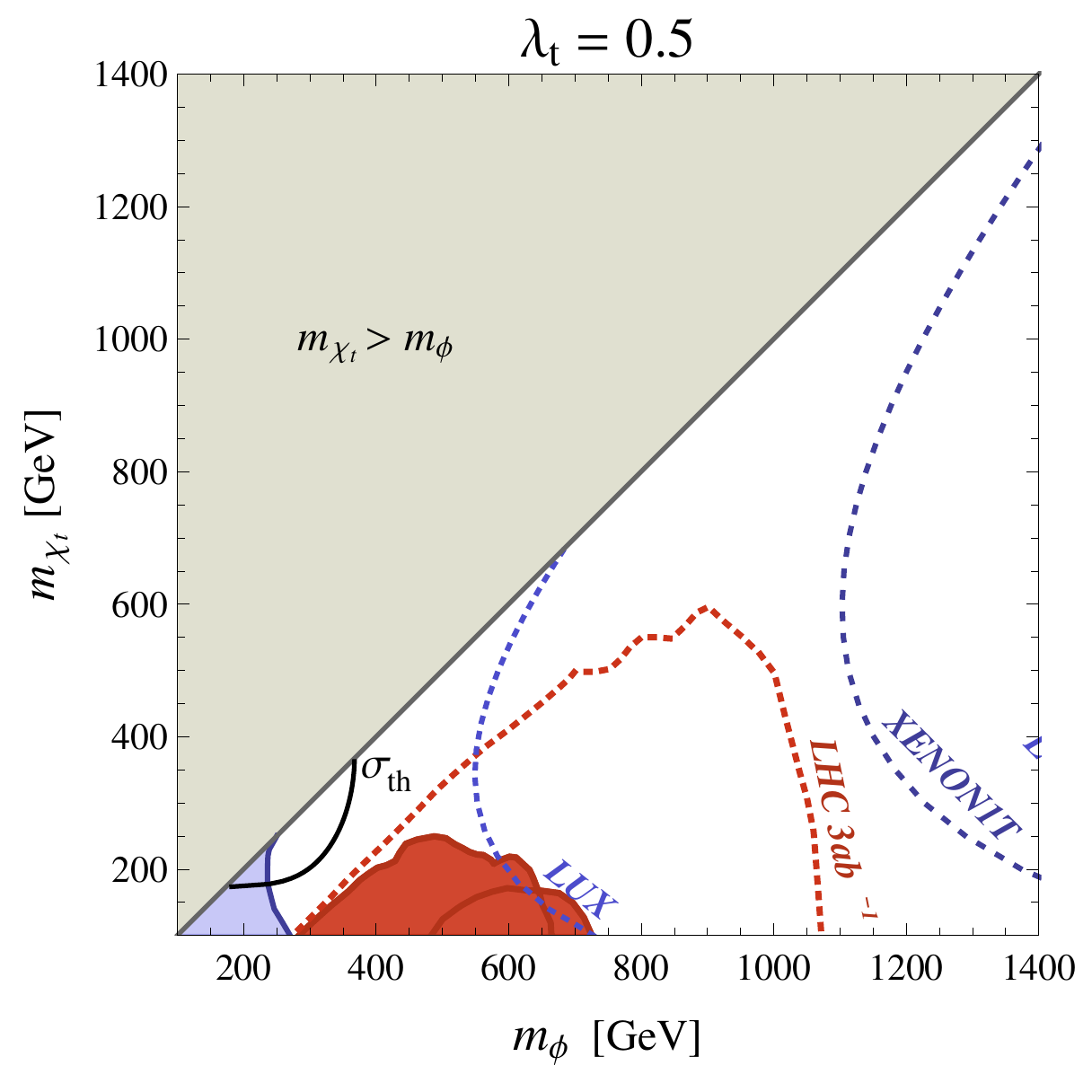}
\includegraphics[width=0.47\textwidth]{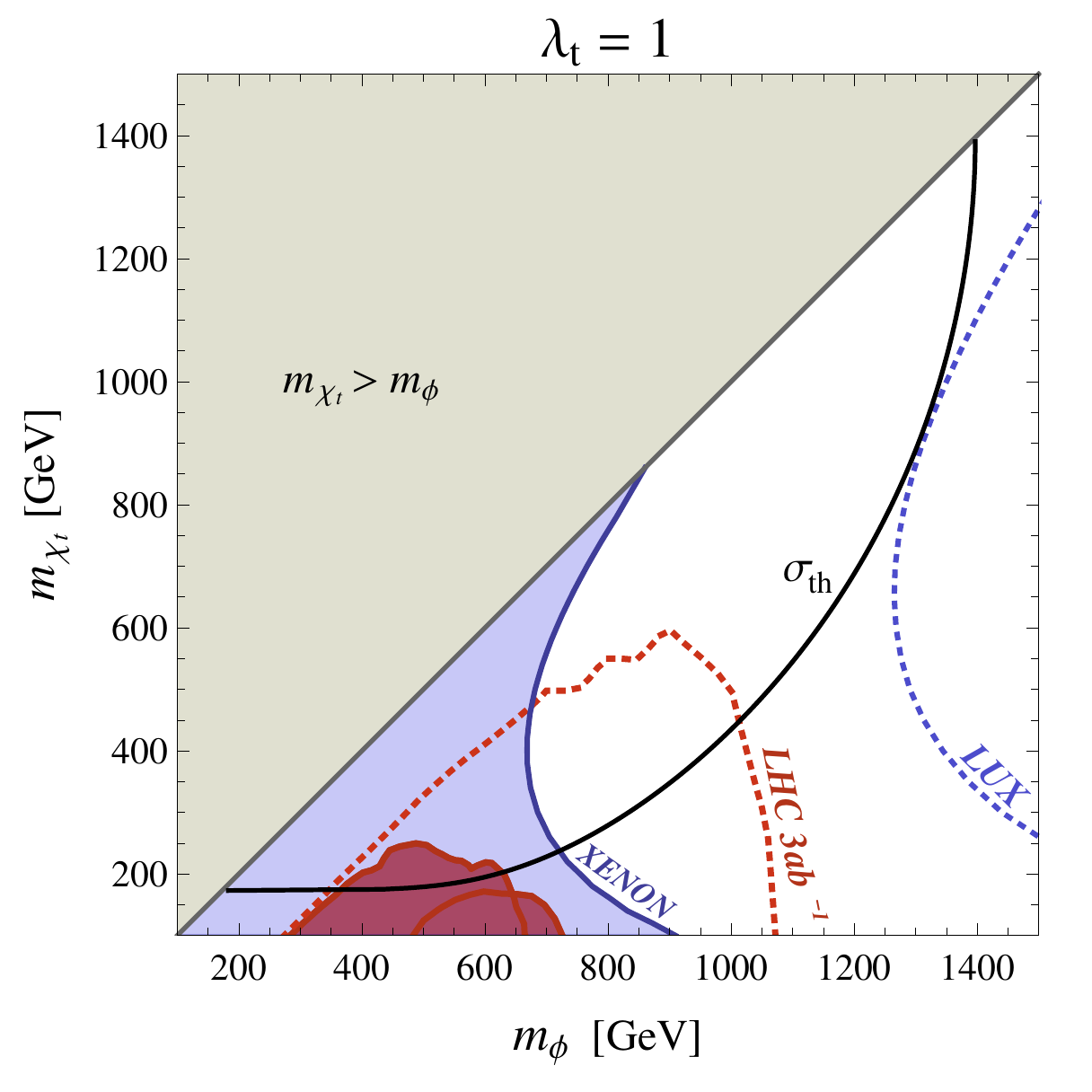}
\caption{Limits on the parameter space of top-flavored DM from LHC
  stop searches with 8 TeV data (solid red area) and XENON100 (solid
  blue area) for $\lambda_t = 0.5$ and $\lambda_t = 1$. The red dashed
  line is a projection for the 95$\%$ CL exclusion limit with a 3000
  $\invfb$ run at the 14 TeV LHC \cite{ATL-PHYS-PUB-2013-002}. For the
  LUX (XENON1T) projection, we show the contour where 1 event is
  expected per 10000 (10$^5$) \kgday. The direct detection limits assume that $\chi_t$ saturates the observed relic density.
  The solid black line labeled $\sigma_{th}$ is where $\sigv = 1.5$ pb, the annihilation cross
  section necessary to obtain a relic density of $\Omega h^2 =
  0.12$. Note that below and to the right of the black line, $\sigv$ is lower and hence
  $\chi_t$ is too abundant, and additional annihilation channels are required.
  \label{fig:lambda}}
\end{center}
\end{figure*}

The parameters needed for $\sigv = 1.5$ pb are shown in
Fig.~\ref{fig:lambda}, given as the curve labeled $\sigma_{th}$. Note
that in the parameter space between this contour and the line $m_\phi
= m_{\chi_t}$, the annihilation cross section is larger and thus
$\chi_t$ cannot be all of the DM. Outside of this curve the cross
section is lower, and other channels are needed so that $\chi_t$ does
not overclose the universe.  In Fig.~\ref{fig:lambda} we have included
only the contribution from the $t\bar t$ mode (\ref{anntop}), which is
valid in the limit of heavy fermionic mediator $m_\psi \gg m_\phi$.
 
\paragraph{\bf Direct Detection}   

\begin{figure}
\begin{center}
\includegraphics[width=0.46\textwidth]{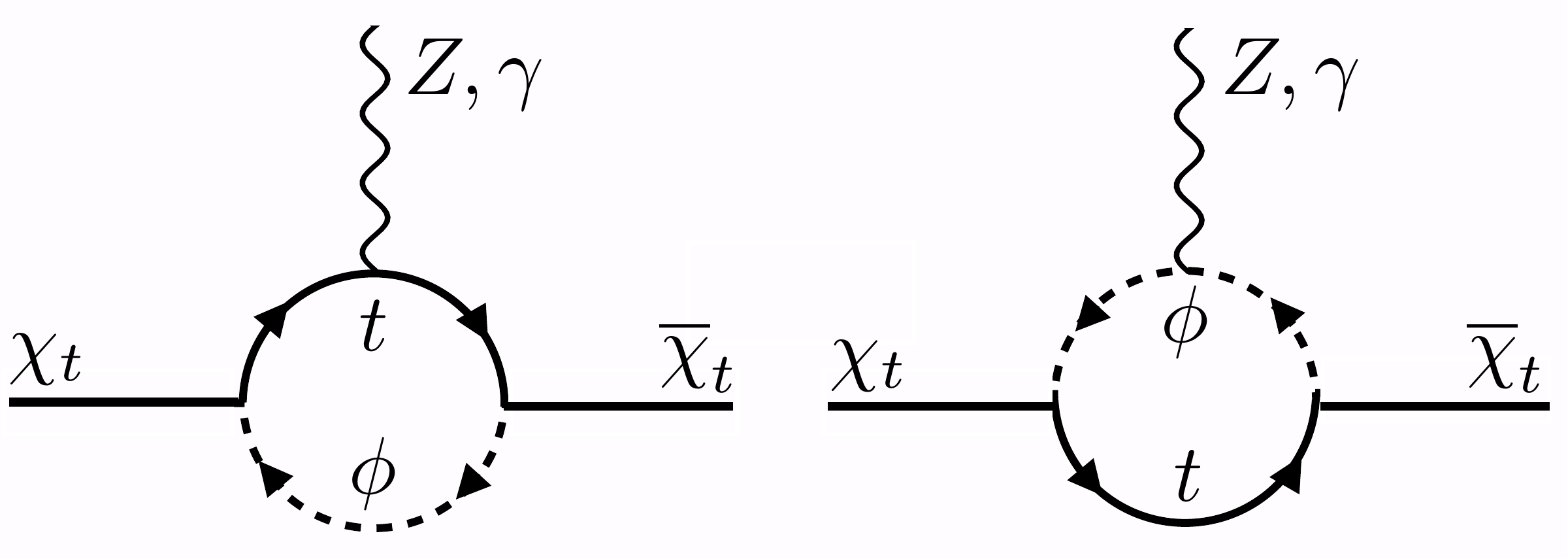}
\caption{One-loop diagrams generating effective $Z,\gamma$ couplings
  for top-flavored DM, with amplitudes given in
  Eqs.~(\ref{eq:Zcoupling},\ref{eq:dipolemoment}).
  \label{fig:DD}}
\end{center}
\end{figure}

The strongest constraints on this model come from direct detection
experiments. Although there is no tree-level coupling of DM with
nucleons, a sizable interaction is generated at one-loop. $Z$ exchange
is the most important process, arising from an effective coupling of
DM to the $Z$. The diagrams shown in Fig.~\ref{fig:DD} generate the
operator
\begin{align}
  {\cal L} \supset g_{Z} Z_\mu \bar \chi_t \gamma^\mu P_L \chi_t
\label{eq:Zcoupling}
\end{align}
at one-loop. We include only the contributions from one-loop $t-\phi$
exchange in our computations, which is valid in the regime $m_\psi \gg
m_\phi$. There are additional diagrams with $\tilde t_i$ and $\psi$ in
the loop that can be numerically important if $\phi$ and $\psi$ have
comparable masses.

The full expression for $g_{Z}$ can be found in the Appendix. In the
limit of $m_\phi \gg m_t, m_{\chi_t}$,
\begin{equation}
  g_{Z} \simeq -\frac{g}{c_w} \frac{\lambda_t^2 N_c}{16 \pi^2} \left(\frac{m_t}{m_\phi} \right)^2 \left( 1 + \log \left[ \frac{m_t^2}{m_\phi^2} \right] \right).
\end{equation}
In general $g_{Z}$ is suppressed by the mass of the fermion in the
loop, but here the large top mass and ${\cal O}(1)$ DM - top coupling
$\lambda_t$ assist in generating a relatively large $Z$ coupling.

The $Z$ coupling mediates a spin-independent (SI) scattering of dark
matter and nuclei. The differential cross section is
\begin{equation}
  \frac{d\sigma_Z}{dE_R} = \frac{m_N}{\pi v^2} 
         \frac{\left( f_p Z +(A-Z)f_n \right)^2}{2} F^2[E_R],
\label{eq:Zrate}
\end{equation}
where the couplings $f_n = g_{Z} G_F c_w/\sqrt{2}g$ and $f_p =
-(1-4s_w^2) g_{Z} G_F c_w/\sqrt{2}g$. Note that we focus on
spin-independent (SI) interactions; although this operator give rise to
spin-independent and spin-dependent scattering with similar cross
sections, the experimental limits for SI interactions are much
stronger.

It is also necessary to consider the same diagrams with an external
photon instead of a $Z$ boson, as shown in Fig.~\ref{fig:DD}.
These generate a magnetic dipole moment for the DM:
\begin{equation}
  {\cal L} \supset  \frac{\mu_\chi }{2} \bar {\chi_t} \sigma^{\mu\nu} {\chi_t} F_{\mu\nu}.
\label{eq:dipolemoment}
\end{equation}
The dipole moment in the limit $m_\phi  \gg m_t, m_{\chi_t}$ is
\begin{equation}
  \mu_{\chi} \simeq \frac{e \lambda_t^2 m_{\chi_t} }{32\pi^2 m_\phi^2},
\end{equation}
with the full one-loop result given in the Appendix. 

We find that magnetic dipole interactions provide a non-negligible
contribution to the rate for larger DM mass, where the dominant
contribution is through the dipole-charge interaction for Xenon. The
scattering cross section is
\begin{align}
  \frac{d\sigma_{DZ}}{dE_R} = \frac{m_N}{\pi v^2} \frac{e^2Z^2}{4} \frac{\mu_\chi^2}{m_N^2} \left( \frac{m_N v^2}{2E_R} - \frac{m_{\chi_t} + 2 m_N}{2 m_{\chi_t}} \right) F^2[E_R].
  \label{eq:dipolecharge}
\end{align}
For $m_{\chi_t} \lesssim m_\phi \sim 1.5$ TeV, the dipole contribution affects
the rate by up to 40-50$\%$, depending on the energy range considered.

Other interactions are also present at one-loop but negligible
compared to $Z$ exchange and dipole-interactions, as we discuss in the
Appendix.

Current experimental limits from XENON100 \cite{Aprile:2012nq}
constrain the parameter space for couplings $\lambda_t \gtrsim
0.5$. (Recall also for $\lambda_t~<~0.35$, other annihilation channels
must be present or the DM will overclose the universe if it is a
thermal relic.)  With increased couplings, larger masses $m_{\chi_t}$
and $m_\phi$ are required to satisfy these limits.

To calculate the limits for XENON100, we compute both $Z$ and dipole
mediated scattering rates over the stated nuclear recoil energy range
of 6.6-30.5 \keVr. The published limits are stated in terms of a
90$\%$ CL exclusion limit on spin-independent nucleon scattering cross
section $\sigma_n$, which apply for $Z$ exchange. However, those
limits are not directly applicable when dipole interactions are
included because of the difference in energy dependence, as can be
seen in Eq.~\ref{eq:dipolecharge}.

Instead, we match the onto the published limits by calculating the
number of expected events from $Z$ exchange only, finding that the
published limit is well approximated by requiring $\le 2$ events with
2324 \kgday. We then include the dipole interactions in the event rate
and re-evaluate the limit. The end result only changes the limits in
$m_{\chi_t}$ and $m_\phi$ by less than $5\%$, so we do not try to model
the energy-dependent acceptance of the experiment.

We also consider projected limits from LUX \cite{2013NIMPA.704..111A},
for which results are expected in the near future.  The expected
energy range is 5-25 \keVr, and with 10000\ \kgday\ exposure. The
procedure for computing these limits is the same as described
above. Again the effect of including dipole interactions only changes
the limits by $\lesssim 5\%$. Finally, we include XENON1T
\cite{Aprile:2012zx} projections, assuming the same energy range as
XENON100 and $10^5$ \kgday\ exposure.  For all of the above
calculations we assume a Standard Halo Model with $\rho_\chi =
0.4\ \GeV$/cm$^3$, $v_{esc} = 550$ km/s, $v_e = 240$ km/s.

We show these results in Fig.~\ref{fig:lambda}. The anticipated LUX
reach can effectively test this model in the case that $\chi_t$ is a
thermal relic annihilating primarily to tops.  The limit and
projection curves approximately follow lines of constant
$g_Z^2/m_{\chi_t}$, which keep the event rate constant.  At fixed
$m_\phi$, the constraint becomes stronger with smaller DM mass
primarily because the rate scales as $1/m_{\chi_t}$, while it again
becomes stronger near $m_{\chi_t} \lesssim m_\phi$ due to the
enhancement of $g_Z$.

\paragraph{\bf LHC signatures}

The collider signatures of this scenario at the LHC depend in detail
on the spectrum of the SM superpartners and the SFDM sector.  An
example spectrum is illustrated in Fig.~\ref{fig:spectrum}.  We will
focus here on the case of a stop LSP.  Provided the stop is lighter
than the DM $\chi_t$ (more specifically $m_{\tilde t_1} <
m_{\chi_t}+m_{\psi}$), then it will decay via the baryon number
violating vertex (\ref{Wbaryon}):
\begin{equation}
\tilde t_1 \rightarrow \bar b \bar s. 
\label{stopdecay}
\end{equation}
Stops decaying via (\ref{stopdecay}) are not constrained by existing
searches for paired dijet
resonances~\cite{Aad:2011yh,ATLAS:2012ds,Chatrchyan:2013izb,Aaltonen:2013hya}.

The DM $\chi_t$ is of course stable and will lead to $\not \!\! E_T$
if produced. There are several ways in which $\chi_t$ can be produced,
as we now discuss. For example, the scalar mediator $\phi$, being
colored, can be pair produced through strong interactions.  Once
produced, it will decay via
\begin{equation}
\phi \rightarrow t \, \bar \chi,  \\
\label{phidecay}
\end{equation}
resulting in a signature of $t \bar t \, +\! \not \!\! E_T$ for $\phi$
pair production.  Remarkably, $\phi$ acts as a ``fake stop'', in that
the signature of $\phi$ pair production mimics exactly the signature
of stop pair production in standard RPC scenarios when the stop decays
to a top and a stable neutralino LSP. The limits (and projected reach)
on direct stop pair production from CMS \cite{Chatrchyan:2013xna} and
ATLAS\cite{ATLAS-CONF-2013-024,ATLAS-CONF-2013-037} can thus be
directly applied to the case of $\phi$ production and are displayed in
Fig.~\ref{fig:lambda}.

Another channel which is important is pair production the fermionic
mediator $\psi$, which will decay via
\begin{equation}
\psi \rightarrow \tilde t_i \bar \chi.    \\
\label{psidecay}
\end{equation}
If the LSP is mostly right handed stop, $\tilde t_1 \sim \tilde t_R$,
then $\psi$ will decay primarily to the stop LSP and the DM $\chi_t$.
The stop LSP will subsequently decay according to
Eq.~(\ref{stopdecay}), leading to a signature of $4 j\,+\!\! \not \!\!
E_T$ for $\psi$ pair production. This signature is similar, though not
identical, to gluino pair production followed by $\tilde g \rightarrow
q \bar q \chi^0$ mediated by a heavy off-shell squark in standard RPC
scenarios.  The gluino searches using data from the 7 TeV run provide
the strongest limits on $\psi$ pair production. This is because the
recent 8 TeV analyses are optimized for a high mass gluino, which,
being a color octet, has a much larger production cross section then
the color triplet $\psi$ in our scenario. The ATLAS
search~\cite{Aad:2012fqa} probes $m_\psi \lesssim 350$ GeV for
$m_{\chi_t} \sim 200$ GeV, $m_{\tilde t_1} \lesssim 100$ GeV. If the
DM $\chi_t$ is lighter, then the sensitivity can be extended up to
higher masses (of order 800 GeV for massless $\chi_t$), although in
this range new annihilation channels are required to obtain a viable
cosmology.

If the new states in the flavored dark sector lie below the MSSM
superpartners, particularly the gluino and first and second generation
squarks, additional signatures are possible. For example, if
kinematically allowed the gluino can decay with a sizable branching
ratio via
\begin{equation}
\tilde g \rightarrow \phi \bar \psi.
\end{equation}
The mediators $\phi$ and $\psi$ subsequently decay according to
Eqs.~(\ref{phidecay}) and (\ref{psidecay}), leading to a multi-top,
multi-jet + $\not \!\! E_T$ final state signature of gluino pair
production.
 
Through their coupling to the top and stops the new states $\chi_t$, $\phi$, and $\psi$ provide 
a relevant contribution to the Higgs mass at two loops. In particular, for the case of a thermal relic $\chi_t$ considered here, 
the required $O(1)$ coupling $\lambda_t$ implies that the new states $\chi_t$, $\phi$, and $\psi$ cannot be 
too heavy without a significant tuning of the weak scale. 
The level of tuning induced by these states is similar to that of the gluino, which also contributes to the Higgs mass at two loops. 
However, in comparison to the gluino in the MSSM with RPC, the LHC limits on the $\phi$ and $\psi$ are generically much weaker, as discussed above. 

\section{Outlook}
\label{sec:outlook}

Supersymmetry with R-parity violation is only weakly constrained by
searches at the LHC, particularly if the LSP decays to jets through
the $\bar u \bar d \bar d$ operator. However, with RPV one abandons a
symmetry rationale for the proton stability and the WIMP miracle.
Minimal Flavor Violation can provide an explanation for the
suppression of proton decay and, as we have demonstrated in this
paper, the stability of WIMP dark matter. This framework provides a
compelling explanation for the naturalness of the weak scale, dark
matter, and the lack of evidence for new physics at the LHC.

In this work we have established the existence of a $Z_3$ discrete
symmetry, {\it flavor triality}, which is a consequence of the MFV
hypothesis. Flavor triality presents the prospect of flavored dark
matter.  We have investigated general aspects of theories of flavored
dark matter in the MFV SUSY framework.  Flavor splittings in the
masses and couplings, which are relevant for cosmology and
phenomenology, arise from non-holomorphic terms in the K$\ddot {\rm
  a}$hler potential. The SFDM framework is compatible with gauge
coupling unification and the mass scale of the dark sector states can
naturally be tied to the weak scale through SUSY breaking.

We have studied in detail a specific scenario of top flavored dark
matter. The dark matter is a thermal relic, annihilating to top quarks
in the early universe. This model can be tested by LUX and future ton
scale Xenon experiments. Furthermore, the scalar mediator $\phi$ can
be directly produced at the LHC and can mimic a stop from standard RPC
scenarios. Therefore, in this scenario it is conceivable that we will
first discover the ``fake stop'' $\phi$ in searches for $t \bar t +
\!\!\not \!\! E_T$, while the true stop responsible for canceling the
Higgs mass quadratic divergences is buried in the QCD multi-jet
background. Many other novel signatures are possible in this scenario,
as we have discussed above.

We have explored just one possible incarnation of SFDM; there are many
other models that would be worthwhile to explore. The condition in
Eq.~(\ref{eq:cond}) ensuring dark matter stability is satisfied for a
variety of representations of $G_q$. There are additional
possibilities in the choice of SM gauge quantum numbers for the dark
matter multiplet $X$ (the only requirement being that it contains a
color and electrically neutral component), as well as the flavor and
gauge representations of the mediator field. Within a given model, 
there are also several candidates for the dark matter particle. For instance in the model
studied in this paper, one could also consider the scalar component as the dark matter,
as well as, {\it e.g.}, up-flavored dark matter, as has been studied recently in simplified DM 
models~\cite{Chang:2013oia},\cite{An:2013xka},\cite{Bai:2013iqa},\cite{DiFranzo:2013vra}.
A systematic investigation of these theories should be carried out.

\subsubsection*{\bf Acknowledgements}
We thank P. Agrawal, R.~Hill, M.~Solon, and D.~Stolarski for helpful discussions. 
This work was supported in part by the Kavli Institute for Cosmological Physics at the 
University of Chicago through grant NSF PHY-1125897 and an 
endowment from the Kavli Foundation and its founder Fred Kavli.
B.B. and L.T.W. are supported by the NSF under grant PHY-0756966 and
the DOE Early Career Award under grant DE-SC0003930.  B.B. and
L.T.W. thank the Aspen Center for Physics and the KITP, Santa Barbara,
where part of this work was completed. BB also 
thanks the INFO 13 workshop, sponsored by the Los Alamos National Laboratory, where part of this work was completed.

\begin{appendix}
\section{Appendix --- Direct detection}
\label{app:directdetection}

This appendix summarizes one-loop amplitudes relevant to direct
detection for the interactions in Eq.~(\ref{eq:interactions}). 
Useful related discussions can be found in Refs.~\cite{Agrawal:2011ze},\cite{Kumar:2013hfa}. 
As discussed in the text, there is a $Z$ coupling,
Eq.~(\ref{eq:Zcoupling}). In the limit of zero momentum transfer ($q^2
\rightarrow 0$) the coupling is
\begin{align}
  g_{Z} &=  \frac{\lambda_t^2 N_c}{16 \pi^2}  \int_0^1 \!\! dy \ (1-y) I(y),
      \label{eq:Zloop}  \\
  I(y) &= g_R \left[  \log \left( \frac{ \deltas }{ \deltaf } \right) - 1 
     - \frac{r_\chi y(1-y)}{\deltas} \right] 
     + g_L  \left( \frac{r_t}{\deltaf} \right), \nn
\end{align}
where $g_{L,R}$ are the couplings of the left- and right-handed top to
$Z$.  The other dimensionless quantities are
\begin{align}
  \deltaf & \equiv   y(1 + r_\chi (y-1)) - r_t(y-1), \\
  \deltas & \equiv    y(r_t + r_\chi (y-1)) + (1-y), \\
  r_t & \equiv \left( \frac{m_t}{m_\phi} \right)^2,  \ 
    r_\chi \equiv \left( \frac{m_{\chi_t}}{m_\phi} \right)^2,
\end{align}
with $\deltaf$ ($\deltas$) corresponding to the propagator factor for
the diagram with emission of the $Z$ from the fermion (scalar) in the
loop.

For an external photon, the diagrams and the calculation are the same,
but with $g_L=g_R=2e/3$. The vector and axial-vector couplings to the
photon vanish in the limit $q^2 \rightarrow 0$, as required by electromagnetic gauge invariance.
There is however a nonzero magnetic dipole moment, $(\mu_\chi/2)
\bar \chi \sigma^{\mu\nu}\chi F_{\mu\nu}$:
\begin{align}
  \mu_\chi &= \frac{e \lambda_t^2 m_{\chi_t}}{32\pi^2 m_\phi^2} \int_0^1 \!\! dy \
  \frac{ 2 y(1-y)^2(1 + r_t + 2  r_\chi y(y-1))}{\deltaf \deltas}.
\end{align}

For reference we also give the amplitude for the (subdominant)
charge-charge contribution to direct detection. Although the vector
coupling to photons is zero in the $q^2=0$ limit (the DM is neutral),
there are $q^2$ corrections that give rise to charge-charge
interactions (as well as velocity-suppressed charge-dipole
interactions). The amplitude has a contribution $b_{\chi}
q^2 \bar \chi \gamma^\mu \chi$, where the coefficient is:
\begin{align}
  b_{\chi} = & \frac{ e \lambda_t^2 N_c }{48 \pi^2 m_\phi^2} \int dy \ I(y),  \\
  I(y) = & \frac{ (1-y)^3}{6} \bigg( \frac{1}{ \deltaf } 
      + \frac{r_t + r_\chi y^2 }{ \deltaf^2 } \nn \\
    &  + \frac{ 2 r_\chi (y-1)y }{ \deltas^2   }  
      + \frac{(2y-1)(r_t - 1)}{ \deltaf \deltas  } \bigg). \nn
\end{align}
The differential cross section is
\begin{equation}
  \frac{d\sigma}{dE_R} = \frac{m_N}{2\pi v^2}  Z^2e^2 b_{\chi}^2 F^2[E_R].
\end{equation}

Other interactions are present but negligible. The coupling of
$\chi_t$ to the Higgs is not important for direct detection because
the coupling of the Higgs to nucleons is small.  There is a scalar
coupling to gluons, but this is only generated at subleading order in
$1/m_\phi^2$ and in the limit of large $m_\phi^2$ the contribution to
direct detection is parametrically suppressed relative to $Z$-exchange
by $(m_{\chi_t} m_n/m_\phi^2)^2$ \cite{Hisano:2010ct}.  A box diagram
with $\chi_u$ in the loop can also generate a 4-fermion operator
coupling $\chi_t$ to $u$ quarks, but the amplitude is suppressed by 
$(\lambda_u)^2$ and the mass of $\chi_u$.

\end{appendix}

\end{document}